\documentstyle[aps,prl,multicol,epsf]{revtex}
\epsfclipon

\begin{document}
\draft

\title{La$_{7/8}$Sr$_{1/8}$MnO$_3$:
long period orbital order with hole stripes}

\author{Michael Korotin$^1$, Takeo Fujiwara$^2$, and Vladimir Anisimov$^1$}

\address{$^1$Institute of Metal Physics, Russian Academy of Sciences, 620219
Ekaterinburg GSP-170, Russia\\
$^2$Department of Applied Physics, University of Tokyo, Bunkyo-ku,
Tokyo 113-8656, Japan} 

%\date{\today}
\date{December 27, 1999}
\maketitle 

\begin{abstract}

The electronic structure of La$_{7/8}$Sr$_{1/8}$MnO$_3$ 
was studied by using the rotation-invariant LSDA~(Local Spin Density
Approximation)+U formalism.
We found the specific kind of charge and orbital order 
for ferromagnetic insulating solution in 
2{\it a}$\times$2{\it b}$\times$4{\it c} 
supercell, where {\it a}, {\it b}, and {\it c} denote cubic perovskite axes. 
The holes are distributed over 3-line stripes in
every second {\it ac} plane along {\it a}~axis.
The stabilization mechanism of the hole stripe is attributed to the
strong $p$-$d$ hybridization.

\end{abstract} 


\begin{multicols}{2}
\narrowtext

%--------- experiments ----------%
Much attention has been focused on doped perovskite
manganite systems La$_{1-x}$Sr$_{x}$MnO$_3$, because of 
its metal-insulator transition, colossal magnetoresistance, spin
structures and properties, and  charge and orbital order. 
The lightly doped La$_{7/8}$Sr$_{1/8}$MnO$_3$ shows the reentrant 
structural transition of orthorhombic structure 
O$^* \rightarrow$O$^\prime \rightarrow$O$^*$, where 
the transition temperatures are 160~K and 260~K, respectively~\cite{Kawano96}.
Further, the sample shows the ferromagnetic transition at
$T_c$=200~K~\cite{Kawano96}.
The ferromagnetic order suppresses drastically the static
Jahn-Teller distortion. 
The resistivity in the O$^\prime$ phase below $T_c$ shows the metallic
temperature dependence and, therefore,
the boundary of the low temperature O$^*$  and O$^\prime$ phases coincides 
with the insulator-metal phase boundary. 
The paramagnetic state is insulating.
Yamada {\it et al.}~\cite{Yamada96} reported the observation of satellite 
spots in the low
temperature O$^*$ phase, corresponding to 
2{\it a}$\times$2{\it b}$\times$4{\it c}
supercell structure~\cite{cubicaxis}, 
and proposed the polaron order in this phase.
Endoh {\it et al.}~\cite{Endoh99} confirmed the superlattice structure
in the low temperature O$^*$ phase of La$_{0.88}$Sr$_{0.12}$MnO$_3$ 
but reported no signal for charge order in the x-ray study. 
They also reported the G-type orbital order 
in the low temperature O$^*$ phase, deduced from the resonant x-ray scattering.
Mizokawa {\it et al.}~\cite{Mizokawa99} found several stable spin, 
orbital and charge
ordered states, using the model Hartree-Fock Hamiltonian. 
They introduced small Jahn-Teller distortion for opening the  band gap. 
However the most striking feature of La$_{7/8}$Sr$_{1/8}$MnO$_3$
may be the observation of orbital order and 
insulating property in a ferromagnetic system without any (or very 
small) Jahn-Teller distortion. 

In this Letter, we present the results of {\it ab-initio} 
self-consistent calculations of the electronic structure of 
La$_{7/8}$Sr$_{1/8}$MnO$_3$, obtained by using the LSDA+U 
formalism~\cite{LSDA+U} in the framework of 
linear muffin-tin orbital method in the atomic sphere 
approximation (LMTO-ASA)~\cite{LMTO}, and the experimentally 
determined crystalline data of low temperature O$^*$
phase~\cite{Kajimoto-Master}. 
The Coulomb and the exchange parameters are 
set to be $U=8.0$~eV and $J=0.88$~eV for Mn $d$- and 
$U=6.0$~eV for O $p$-shell, respectively~\cite{UandJ}.
The inclusion of LSDA+U potential correction for
O $p$-shell was found to be crucial
in the present calculations, since
it controls the value of charge transfer energy 
between Mn $d$- and O $p$-valence orbitals~\cite{SIC}.

%--------- undoped  ----------%

\begin{figure} 
\vspace{-1.5cm}
\centerline{ 
%              \epsfile{file=LSMOfig1i.eps,width=4.5cm} 
%              \hspace{5mm}
%              \epsfile{file=LSMOfig1ii.eps,width=4.5cm}
%              \hspace{5mm}
%              \epsfile{file=LSMOfig1iii.eps,width=4.5cm}  
              \epsfxsize=4.5cm \epsffile{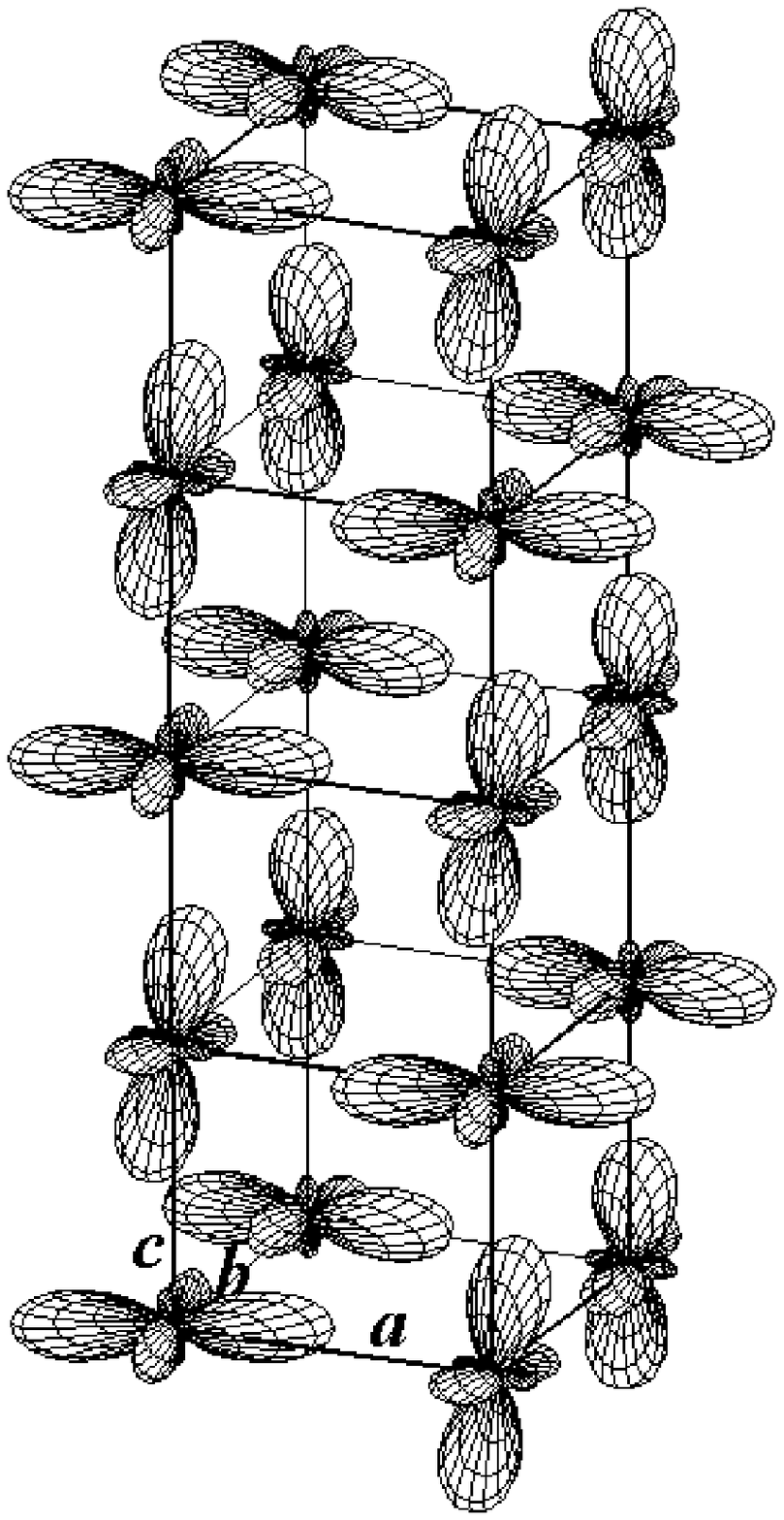} 
              \hspace{5mm}
              \epsfxsize=4.5cm \epsffile{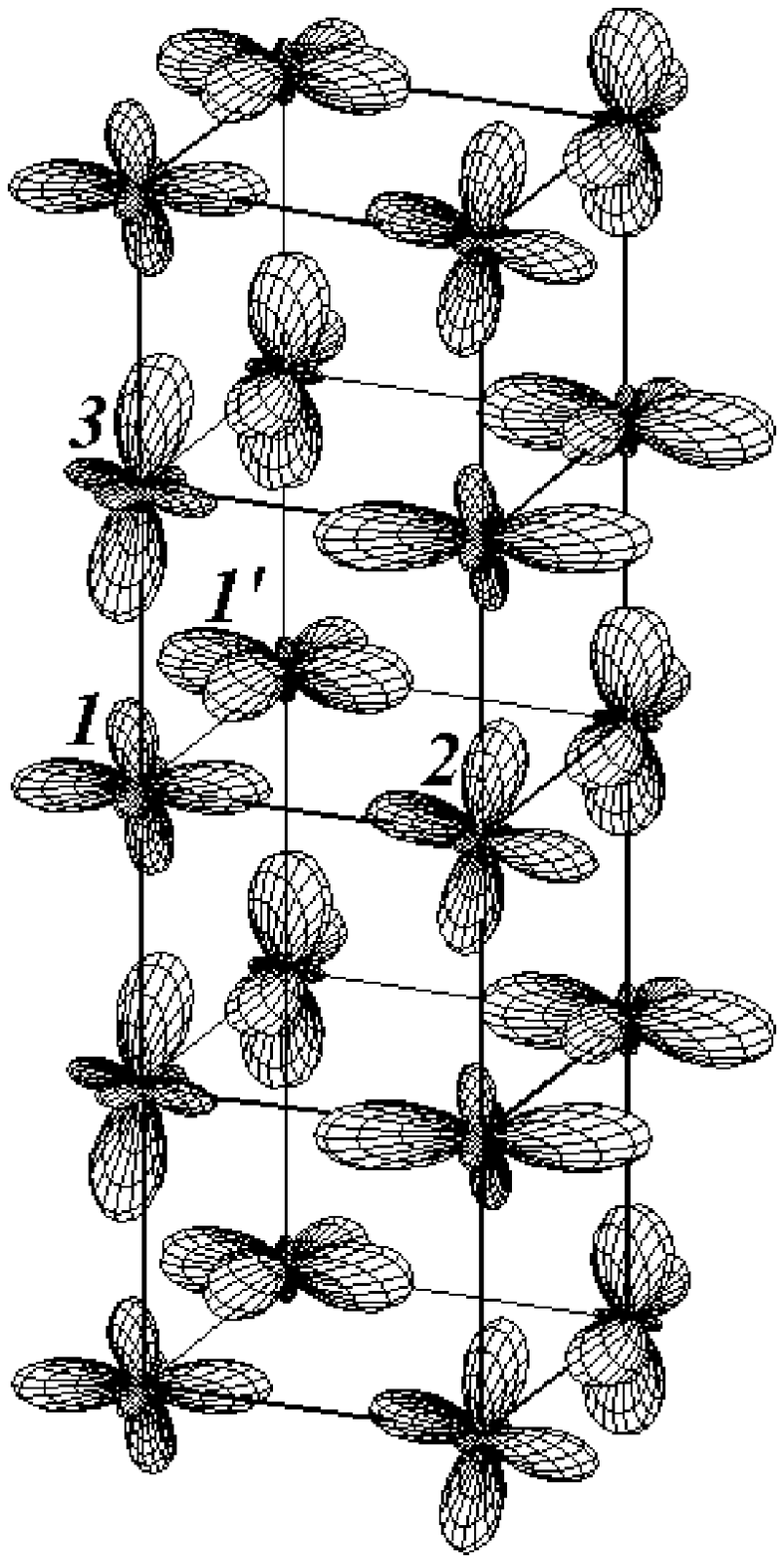} 
              \hspace{5mm}
              \epsfxsize=4.5cm \epsffile{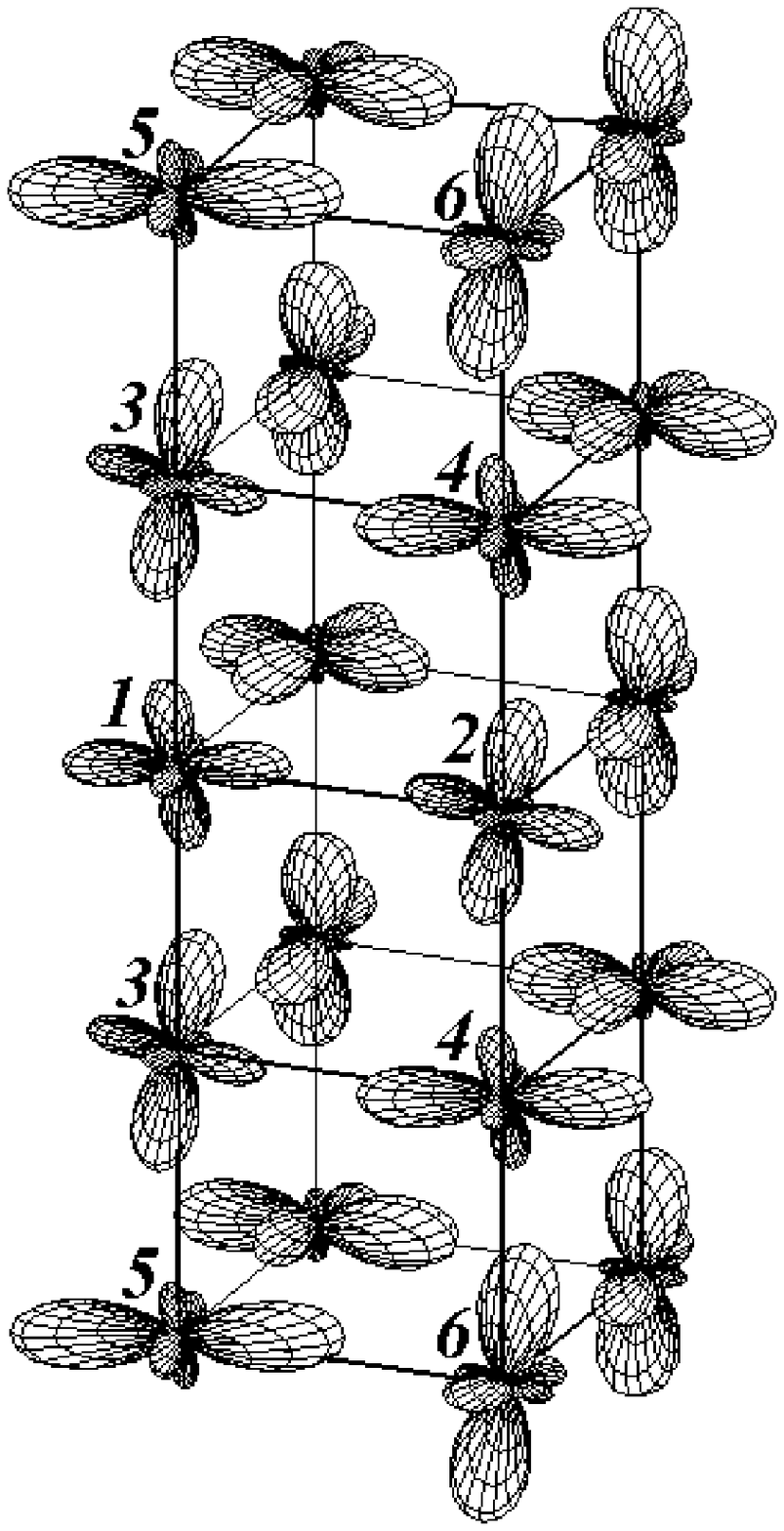} 
              }	           
{\large \bf  \hspace{2.4cm} (i) \hspace{4.6cm} (ii) \hspace{4.2cm} (iii) }
\vspace{0.3cm}
\caption {Angular distributions of e$_g$ electron spin density (ADESD)
of Mn atoms and corresponding spatial orbital order in the 
$2a\times2b\times4c$ unit cell of 
La$_{7/8}$Sr$_{1/8}$MnO$_3$ compound for
undoped~(i), doped metallic~(ii), and doped insulating~(iii) 
solutions.  For case (iii) there are translated atoms from the top to 
the bottom plane along {\it c} axis, and there are no translated atoms 
along {\it a} and {\it b} axes. The unit cells for (i) and (ii) 
cases are increased up to the same size as for (iii) case to make the 
comparison more convenient. The ADESD is defined as
$\rho(\theta,\phi)=\sum_{m,m'}Q_{m,m'}Y_m(\theta,\phi)Y_{m'}(\theta,\phi)$,
where $Q_{m,m'}=n^{\uparrow}_{m,m'}-n^{\downarrow}_{m,m'}$
is 2$\times$2 e$_g$-orbital occupation matrix obtained in our
self-consistent calculation and $Y_m(\theta,\phi)$ are corresponding
spherical harmonics.}
\label{orbitals} 
\end{figure}

\begin {figure} 
\vspace{-0.6cm}
\centerline{ 
%             \epsfile{file=LSMOfig2.eps,width=8.0cm} 
            \epsfxsize=8.0cm \epsffile{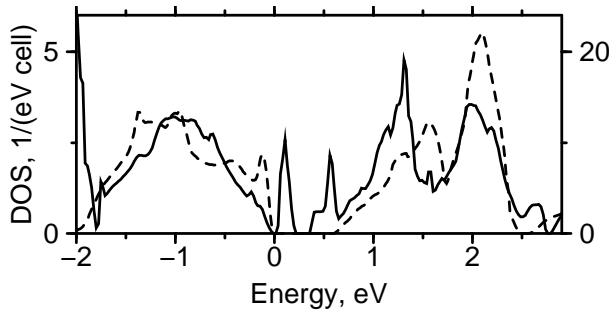} 
           } 
\vspace{0.15cm}
\caption {Total density of states in the vicinity of Fermi energy
($E_F=0$) for doped insulating
La$_{7/8}$Sr$_{1/8}$MnO$_3$ (solid line; right ordinate) 
and model undoped LaMnO$_3$
(dashed line; left ordinate) compounds.}
\label{dos} 
\end{figure}

In order to understand the nature of the doped material, 
we have calculated the
electronic structure of the model ferromagnetic undoped LaMnO$_3$
compound with the 
crystal structure parameters corresponding to the doped case. 
It was found that one e$_g$ majority spin electron per Mn site
occupies alternatively
$\frac{1}{\sqrt{2}}(\phi_{3z^2-r^2}+\phi_{x^2-y^2})$
and $\frac{1}{\sqrt{2}}(\phi_{3z^2-r^2}-\phi_{x^2-y^2})$
orbitals~\cite{lcs,newTF1} (Fig.\ref{orbitals}~(i)).
The band gap of 0.58~eV is opened in the energy spectrum
between majority spin states~(Fig.\ref{dos}), since there is already 
an extended gap of 4.42~eV in the minority spin subband in the 
vicinity of the Fermi energy.  
One can suppose that unusual ferromagnetic insulating state of model
undoped LaMnO$_3$ compound is stabilized 
due to fine balance between double- and super-exchange interactions.
Without orbital order the double-exchange mechanism would bring the 
ferromagnetic metal ground state.
The insulating state appears due to the orbital
polarization in the twofold degenerated e$_g$ majority spin band
occupied by one electron. 
The super-exchange interaction in orbitally ordered state
is ferromagnetic due to specific overlapping of 
$\frac{1}{\sqrt{2}}(\phi_{3z^2-r^2}+\phi_{x^2-y^2})$  
and $\frac{1}{\sqrt{2}}(\phi_{3z^2-r^2}-\phi_{x^2-y^2})$ 
orbitals on neighboring Mn ions on {\it ac} plane.
That is an example of an 
{\em antiorbital} but {\em ferromagnetic} order 
(for review of mutual correspondence between orbital order 
and preferable magnetic interactions in insulators containing 
orbitally degenerate transition metal ions 
see, {\it e.g.} \cite{KugelKhomskii}).
The alignment of the same e$_g$ orbitals along $b$ direction
(see Fig.~\ref{orbitals}~(i)) gives rise to the larger band width 
and resultant further stability than the case of
the alignment of the alternative e$_g$ orbitals. 

%--------- doped 1  ----------%
To investigate the influence of the  Sr doping,
we have calculated the energy spectrum of
La$_{7/8}$Sr$_{1/8}$MnO$_3$ for the minimal possible unit cell 
containing eight Mn atoms and one hole. It was found to be a ferromagnetic 
half-metal with low density  of states at the Fermi level equal

\end{multicols} 
\widetext

\begin{multicols}{2}
\narrowtext

\noindent
 to
0.10~states/(eV$\cdot$atom Mn).
One would expect the uniform distribution of hole state over all
eight Mn atoms for the metallic solution. However it is not the case. 
The Mn1 atom (for denotations see Fig.~\ref{orbitals} (ii))
shows the maximum loss of its $d$-electron density 
(0.06~$e$) and reduces its $d$-magnetic moment by 0.10~$\mu_B$
in comparison with undoped state~\cite{magnitude}. 
This deficiency of $d$-electron density looks as the
decrease of the e$_g$ orbital size in Fig.~\ref{orbitals} (ii).
Thus all these facts
indicate the preferable occupation of Mn1 site by the hole states. 

The influence of Mn1 hole on the nearest sites
Mn2 and Mn3 (Fig~\ref{orbitals} (ii)) is so strong that the changes
of Mn2 and Mn3 atoms in $d$-occupancy (0.04~$e$), 
$d$-magnetic moment (0.07~$\mu_B$) and e$_g$
orbital density angular redistribution allow to consider these sites
as those containing the hole together with Mn1.
At the same time the change of e$_g$ states of Mn1$^\prime$ atom 
is mainly in shape.
There are no essential changes of $d$-occupancy and $d$-magnetic moment
in comparison with model undoped case 
in the electronic states of second plane Mn atoms
(backplanes in Fig.~\ref{orbitals} (i) and (ii) look similar).
Thus we conclude that one hole per eight Mn sites is distributed
in such a way that only three Mn atoms out of eight
change significantly their e$_g$ electronic shell. 

All these three hole-doped Mn atoms are in the same {\it ac} plane
(frontplane in Fig.~\ref{orbitals} (ii)). In frontplane 
the e$_g$ orbitals of the hole-doped Mn atoms overlap strongly
leading to the formation of a wide band. On this stage -- for small unit cell 
restricted by eight Mn atoms -- double exchange would win super-exchange
and the spectrum is metallic. 
It is interesting to note that partial densities of states
have  vanishing values at the Fermi level for Mn atoms on the backplane. 
Thus, the backplane remains insulating even for doped metallic
solution.

%--------- MK4-bdc  ----------%
However, the real La$_{7/8}$Sr$_{1/8}$MnO$_3$ compound demonstrates 
the insulating properties. We observed the appearance of the small
(0.03~eV) band gap in the energy spectrum of La$_{7/8}$Sr$_{1/8}$MnO$_3$ 
(Fig.~\ref{dos}) when the unit cell was increased up to
2{\it a}$\times$2{\it b}$\times$4{\it c} size, where {\it a}, {\it b}, 
and {\it c} denote cubic perovskite axes.
In this ferromagnetic insulating case there are two holes per sixteen Mn atoms. 

\begin {figure} 
\centerline{  
%             \epsfile{file=LSMOfig3.eps,width=3.2cm} 
             \epsfxsize=3.2cm \epsffile{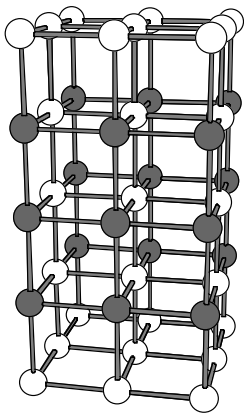} 
           } 
\vspace{2mm}
\caption {The pattern of hole order for
La$_{7/8}$Sr$_{1/8}$MnO$_3$ obtained in our calculations. Gray circles
represent Mn sites containing the hole states, white circles --
Mn$^{3+}$ ions.}

\label{structures} 
\end{figure}

One can see two ``impurity'' peaks in the density of states
in the energy region
of former band gap for model undoped case (Fig.~\ref{dos}). 
The low-energy peak has capacity  precisely
equal to 1 and is well separated from the rest band.  
It is mostly originated from Mn3 and Mn4 states with the
admixture of Mn2 states. 
The second peak is not split enough from the band to
be single but its capacity reaches  1 in the deep. 
States of Mn1 and Mn2 contribute mostly to this  peak.
In these two ``impurity'' peaks,
79~\% of hole states  are localized on six Mn1-Mn4
atoms forming a 3-line stripe (Figs.~\ref{orbitals} (iii), 
\ref{structures}) and their\break

\noindent
nearest oxygen neighbors.
The distribution of hole states between these Mn and O atoms 
belonging to the stripe is 44~\% and 35~\%, correspondingly. 

The same conclusion -- hole states localization on stripe -- 
follows from the values of Mn $d$-occupancy and $d$-magnetic moment
values. The deficiency of $d$-occupancy
is 0.07, 0.05, 0.04, 0.04~$e$ and that of $d$-magnetic moment
is 0.11, 0.09, 0.07, 0.07~$\mu_B$ for Mn1, Mn2 Mn3, and Mn4 atoms,
correspondingly; for the rest of Mn atoms these decreases
are essentially less. 

The comparison of the mutual orientation of e$_g$ orbitals 
between undoped and doped solutions also support this conclusion.
One can see from Table~\ref{coef} for frontplane Mn atoms that
Mn5 and Mn6 atoms contain no holes in their majority spin ($\uparrow$)
e$_g$ shell. Wave function of e$_g^\uparrow$ electron of Mn6 atom is the 
closest to that of undoped case (close to equal contributions of
$\phi_{3z^2-r^2}$ and $\phi_{x^2-y^2}$), whereas e$_g^\uparrow$ 
electron of Mn5 atom occupies the orbital close to 
$\phi_{3y^2-r^2}$ wave function~\cite{lcs}.
Other Mn atoms presented in Table~\ref{coef} contain the hole because of their
reduced e$_g^\uparrow$ occupancy. The contributions of $\phi_{3z^2-r^2}$
and $\phi_{x^2-y^2}$ wave\break  

\vspace{3mm}

\noindent
functions provide the angular distribution presented in 
Fig.~\ref{orbitals} (iii).  And as for the model doped metallic case,
the backplane of Mn atoms  is free of holes. The holes are localized
on Mn atoms in the 3-line stripes along {\it a} axis in every second 
{\it ac} plane (Fig.~\ref{structures})~\cite{independence}.

In LSDA+U formalism, the less occupied is a localized orbital the higher is
its energy position~\cite{LSDA+U}. The difference in e$_g$ orbital
occupancy of the most empty Mn1 and {\it e.g.} Mn2,3 atoms (see
Table~1) explains the appearance of two hole peaks. There is no
contribution to the first low-energy hole peak from Mn1 e$_g$ states.
They are the most empty and contribute only to the second hole peak.

One would expect that two holes per unit cell move apart from each 
other as far as possible to minimize the Coulomb repulsive energy. 
However the situation is more complicated due to the strong 
$p$-$d$ hybridization. With the help of model hamiltonian
describing in the simplest way the interaction $t$ between oxygen $p$ 
level with $d$ levels of two neighboring manganese ions
\begin{displaymath}
%\mathbf{H}=
{\bf H}=
\left( \begin{array}{ccc} 
d_1 & 0 & t \\ 0 & d_2 & t \\ t & t & p \\
\end{array} \right)
\end{displaymath}
it is easy to simulate the process of hole distribution among the
Mn sites. Suppose that one hole is on the first Mn site. Then the 
presence of the hole leads to the
increasing of the $d_1$ energy due to its smaller occupancy and hence 
to the increasing of the $d_2$ level energy due to hybridization via 
$p$ oxygen. Then the second hole will occupy the $d_2$ level because
it has higher energy. If one hole locates on one specific site 
(let it be Mn1 in Fig.~\ref{orbitals} (iii)), that leads to  
decreased electron content on the neighboring sites in {\it a} 
(Mn2) and {\it c} (two Mn3) directions as it was demonstrated for the
doped metallic solution. The second hole occupies Mn2 site with 
higher energy level and in its turn makes
more empty its nearest neighbors along {\it c} direction (two Mn4).
The hole-rich sites Mn1 and Mn2 are not the same due to essentially 
different orientation of e$_g$ orbitals of neighboring Mn atoms along 
{\it c} axis. Such system
consisting of {\it six} hole-doped Mn atoms (Mn1, Mn2, two Mn3, and two Mn4) 
per 2{\it a}$\times$2{\it b}$\times$4{\it c} supercell is closed
and produces small influence on the atoms of next $ab$-slab
(Mn5 and Mn6) and on the Mn atoms in the back $ac$-plane.
The break of long-range e$_g$ orbital\break

\begin{table}
\vspace{-1mm}
\caption{Contributions of e$_g$ symmetry wave functions to the
occupied majority e$_g$ orbital of different frontplane Mn atoms 
in the case of doped insulating La$_{7/8}$Sr$_{1/8}$MnO$_3$.}
\begin{tabular}{cccc}
\multicolumn{1}{c}{atom}&
\multicolumn{1}{c}{$\phi_{3z^2-r^2}$} &
\multicolumn{1}{c}{$\phi_{x^2-y^2}$} &
\multicolumn{1}{c}{occupancy} \\
\hline
\multicolumn{1}{c}{Mn1} &
\multicolumn{1}{c}{.044} &
\multicolumn{1}{c}{$+$.999} &
\multicolumn{1}{c}{.852} \\
\multicolumn{1}{c}{Mn3} &
\multicolumn{1}{c}{.257} &
\multicolumn{1}{c}{$-$.966} &
\multicolumn{1}{c}{.941} \\
\multicolumn{1}{c}{Mn5} &
\multicolumn{1}{c}{.537} &
\multicolumn{1}{c}{$+$.844} &
\multicolumn{1}{c}{1.00} \\
\multicolumn{1}{c}{Mn2} &
\multicolumn{1}{c}{.117} &
\multicolumn{1}{c}{$-$.993} &
\multicolumn{1}{c}{.941} \\
\multicolumn{1}{c}{Mn4} &
\multicolumn{1}{c}{.348} &
\multicolumn{1}{c}{$+$.938} &
\multicolumn{1}{c}{.952} \\
\multicolumn{1}{c}{Mn6} &
\multicolumn{1}{c}{.657} &
\multicolumn{1}{c}{$-$.754} &
\multicolumn{1}{c}{1.00} \\
\end{tabular}
\label{coef}
\end{table}

\noindent
overlapping along {\it c} 
axis leads to the appearance of the insulating state. 

%---- Discussions, comparison with experiments -----------
The calculated energy gap of 0.03~eV in the doped insulator 
is very small and can be smeared out with elevating temperature. 
This temperature may correspond to $T_{OO}$ 
or the ${\rm O}^*\rightarrow {\rm O}^\prime$ transition temperature.  

Present orbital order for one undoped plane looks similar to that
previously proposed in \cite{Endoh99} and thus can be used for the 
explanation of their experimental dependence of polarization on 
azimuthal scan. The fact that 
holes accumulate on one doped $ac$ plane but there are no holes on 
the other plane is consistent with the observed 
superlattice satellite spots~\cite{Yamada96} in neutron diffraction. 
The charge difference between hole sites and other sites is actually 
small. Since two holes are localized on six Mn atoms, the hole charge is
$1/3$ in average. It was demonstrated earlier that hole charge is
also spread over oxygen atoms neighboring the stripe, causing the further
reduction of charge difference between doped and undoped sites. 
This may be the reason why several experiments could not detect 
the present pattern of charge order.

It was demonstrated theoretically for a wide range of magnetic 
compounds containing 
Jahn-Teller ions~\cite{KugelKhomskii} that the exchange mechanism 
(caused by interplay between on-site Coulomb interaction {\it U}
and intersite hopping {\it t}) by itself, without taking Jahn-Teller
distortions into consideration, is able to give a correct picture of the
orbital order. This conclusion was supported by band structure 
calculations~\cite{withTerakura}
of undistorted {\em model}  PrMnO$_3$ compound.
The system La$_{7/8}$Sr$_{1/8}$MnO$_3$  is unique {\em real} compound 
which has an orbital order without Jahn-Teller distortion. 
And the super-exchange mechanism overcomes the double exchange mechanism 
and  provides the insulating state with orbital order.

%------ conclusion --------%
In conclusions, we have investigated the electronic structure 
of ferromagnetic insulating La$_{7/8}$Sr$_{1/8}$MnO$_3$ 
using the LSDA+U formalism. 
We obtained alternative sequence of undoped and doped $ac$ planes 
with alignment of holes in 3-line stripe along  $a$ axis. 
The long period orbital order and holes in stripe  are 
consistent with experimentally observed properties of this compound. 

%%%%%%%%%%%%% Acknowledgments  %%%%%%%%%%%%%%%%%%%%%%%%%%%%%%%%
This work is supported by Grant-in-Aid for COE Research
``Phase Control in Spin-Charge-Photon Coupled Systems'',  
and Grant-in-Aid from the Japan Ministry of Education, Science, Sports 
and Culture. It is supported partly by Russian Foundation for Basic
Research (RFBR Grant No. 98-02-17275). 

%%%%%%%%%%%%% Bibliography  %%%%%%%%%%%%%%%%%%%%%%%%%%%%%%%%%%%%

\end{multicols}


\begin{references}
\bibitem{newTF1}If one calculates the hypothetical ferromagnetic 
state for 
real pure LaMnO$_3$ compound with strong Jahn-Teller distortion, 
then one e$_g$ electron occupies on every Mn site $\phi_{3z^2-r^2}$ orbital
directing to the most remote oxygen atoms. In the present case
when there is no influence of structural factors
the occupied e$_g$ orbitals are different.
\bibitem{independence} There is no dependence of results obtained on 
initial hole configuration. Moreover, the results 
do not depend on the specific position of the Sr impurities.
\end{references}
\end{document}